# MOMENTUM JUMPS IN QUASI-2D BALLISTIC SYSTEMS

D. Dragoman*– Univ. Bucharest, Physics Dept., P.O. Box MG-11, 077125 Bucuresti-Magurele, Romania

ABSTRACT

It is shown that the depopulation of magnetoelectric subbands of ballistic electrons in quasi-2D systems, due to an increased magnetic field parallel to the 2D electron gas plane, produces a momentum jump of the ballistic electrons in a direction transverse to the magnetic field. The present technological achievements allow the observation of this new phenomenon, which can be used to implement qubit states or electron switches.

* Correspondence address: P.O. Box 1-480, 014700 Bucharest, Romania, email: danieladragoman@ yahoo.com

Quasi-2D ballistic systems in the presence of magnetic fields have become testing grounds for a wealth of quantum phenomena. The majority of studies have considered magnetic fields applied perpendicular to the plane of the 2D electron gas (2DEG), configuration in which Shubnikov-de Haas oscillations of the magnetoresistance were observed, magnetic depopulation of electron sublevels were demonstrated and the quantum Hall effect has been evidenced (see for example [1] and the references therein). The case of a magnetic field with magnitude $B$, parallel to the 2DEG plane, has received considerable less attention since it was found that the main effect of the magnetic field in this case was to produce an additional confinement of the already quantized transverse energy levels that leads to a diamagnetic shift of the sublevel separation with increasing $B$ [1-3]. As a result, a magnetic depopulation of the hybrid magnetoelectric subbands occurs and oscillations of the magnetoresistance appear. More interesting phenomena occur in tilted magnetic fields, when the parallel magnetic field component shifts (relative to one another) the occupied sets of Landau levels induced by the perpendicular magnetic field component, and can cause even splitting of the cyclotron resonance when these sets are brought into coincidence [4]. The effect of a parallel magnetic field on quasi-1D systems is studied in Ref.[5].

This paper demonstrates that the consequences of magnetic depopulation of quasi-2D ballistic systems in parallel magnetic fields are enriched by the prediction of a new phenomenon: magnetic depopulation observed when, at increasing parallel magnetic fields, the Fermi energy of electrons passes through different magnetoelectric levels, is accompanied by jumps in electron momentum. These momentum jumps of ballistic electrons, which propagate with scattering in the 2DEG plane, are a consequence of the necessity of conservation of the total electron energy at very low temperatures ($T \cong 0$), when interactions with phonons are practically absent. The momentum jumps can be observed by measuring the current of appropriately placed electrostatic gates, which collect the electrons that deviate from their original direction, and can be used to implement electron switches and, in particular, quantum logic states (qubits).

The theory of quasi-2D ballistic electron propagation is well known and has analytical solutions for a confining parabolic transverse potential $V$. Let us consider a 2DEG in the $(x, y)$ plane, as depicted in Fig.1a, confined by a transverse potential $V(z) = m w_0^2 z^2 / 2$, with $m$ the electron effective mass and $w_0$ a constant with dimensions of frequency, and a magnetic field $B$ applied along the $y$ direction. Neglecting spin splitting and choosing a vector potential of the form $\mathbf{A} = (Bz, 0, 0)$, one finds the Hamiltonian of electron propagation

$$H = \frac{p_x^2}{2M} + \frac{p_y^2}{2m} + \frac{p_z^2}{2m} + \frac{m w^2}{2}(z - \bar{z})^2 \qquad (1)$$

where $M = m(1 + w_c^2 / w_0^2)$, with $w_c = eB/m$ the cyclotron frequency, $w = (w_0^2 + w_c^2)^{1/2}$ and $\bar{z} = (\hbar k_x / eB)(w_c^2 / w^2)$ [1]. The electron has a free-like propagation along $x$ and $y$, characterized by electron wavevector components $k_x$ and $k_y$, respectively, the electron wavefunction being given by

$$\Psi(x, y, z) = \exp(i k_x x + i k_y y) f_n(z - \bar{z}) \qquad (2)$$

where $f_n(z) = (2^n n! \sqrt{\pi} z_0)^{-1/2} \exp[-(z - \bar{z})^2 / 2 z_0^2] H_n[(z - \bar{z}) / z_0]$ are the eigenfunctions of the harmonic oscillator, with $z_0 = (\hbar / m w)^{1/2}$. Thus, the influence of a parallel magnetic field is to modify the effective electron mass along the transverse direction (to $B$) of free propagation and to alter the confinement strength along the other transverse direction. The energy spectrum

$$E_{n,k_x,k_y}(B) = \frac{\hbar^2 k_x^2}{2M} + \frac{\hbar^2 k_y^2}{2m} + \hbar w\left(n + \frac{1}{2}\right) = E_{n,k_x}(B) + \frac{\hbar^2 k_y^2}{2m} \quad (3)$$

has a transverse magnetic-field-dependent component $E_{n,k_x}(B)$ consisting of several subbands labeled by $n$ and characterized by the $B$-dependent electron effective mass $M$.

Each subband is diamagnetically shifted from the $B = 0$ curve by

$$\Delta E_n = E_{n,0}(B) - E_{n,0}(0) = \hbar\left(n + \frac{1}{2}\right)\left(\sqrt{w_0^2 + w_c^2} - w_0\right) > 0. \quad (4)$$

These subbands are occupied by electrons according to Eq.(1) in Ref.[2]; the number of occupied subbands depends on the Fermi energy level of electrons.

For simplicity, let us consider that only two subbands are occupied by electrons at $B = 0$ and that the corresponding subband populations are $N_0$ and $N_1$. The corresponding energy dispersion diagrams are shown in Fig.1b. If a magnetic field is now applied parallel to the 2DEG the subband populations change to [2]

$$N_0(B) = N_0 + (m/2\hbar^2 p)(\Delta E_1 - \Delta E_0), \quad (5a)$$
$$N_1(B) = N_1 + (m/2\hbar^2 p)(\Delta E_0 - \Delta E_1). \quad (5b)$$

Since, according to Eq.(4), $\Delta E_1 > \Delta E_0$ it follows that the higher subband becomes depopulated as the magnetic field increases, and it becomes empty for a magnetic field

$$B_1 = \frac{m}{e}\left[\left(w_0 + \frac{2\hbar p N_1}{m}\right)^2 - w_0^2\right]^{1/2}. \quad (6)$$

At this point, according to the literature, the magnetoresistance drops discontinuously to a lower value.

Let us look in more detail to what happens with the electrons in the higher subband when the parallel magnetic field attains the critical value $B_1$. We have implicitly assumed that the electrons are ballistic, i.e. they propagate without scattering, and that the temperature is $T = 0$ (or close to). Thus, the surplus energy of electrons caused by jumping to an energy level with a lower value cannot be overtaken by phonons, which are absent at $T = 0$ (or their number is extremely small), and therefore at $B_1$ the electrons in the higher subband have only one choice: to "jump" into the lower subband maintaining their total energy. This jump is represented in Fig.1b. This means that if the balistic electrons are initially launched into the 2DEG with $k_x = 0$, i.e. are launched in the direction of the applied magnetic field, their moment must change such that, after the "jump",

$$\frac{\hbar^2 k_x^2}{2M} = \hbar\left(\sqrt{(eB_1/m)^2 + w_0^2} - w_0\right) \neq 0. \tag{7}$$

Thus, the ballistic electrons change abruptly their propagation angle with respect to the direction of the applied magnetic field from 0 to $q \neq 0$, where $\tan q = \pm k_x / k_y$ with $k_x$ the positive solution of Eq.(7). The value of $k_y$ can be determined from $k_y = m\mathbf{m}F/\hbar$, where $\mathbf{m}$ is the electron mobility and $F$ the value of the electric field applied along $y$. Instead of one direction for electron propagation there are now three: the initial middle one and two outer ones, symmetric with respect to the first (see Fig.1a, where the outer electron trajectories are represented with dashed lines and the middle, initial electron trajectory is represented with solid line).

According to the author's knowledge no predictions of such momentum jumps have been reported based either of theory or on experimental results. The remarkable fact is that the jumps in momentum occur while maintaining constant the total electron energy; they are the result of redistribution of energy between different components of electron motion. It is interesting to compare these magnetic-field induced momentum jumps with the magnetic-field-dependent conditions of electron tunneling across a quantum well (a 2DEG) when the magnetic field is parallel to the 2DEG [6]. The jump to a state with $k_x \neq 0$ when the magnetic field reaches the critical value $B_1$ can also be seen as tunneling to allowed electron states.

The prediction of momentum jumps has been supported by calculations based on a parabolic confining potential $V(z) = m\omega_0^2 z^2 / 2$ in the absence of the magnetic field. This choice is supported by the possibility of easily finding analytical formulae; it is also a widely used model in literature. However, in some previous works [2-4] a triangular confining potential has been used to model the experimental results obtained in narrow-gap materials. This asymmetric (in $z$) triangular potential does not change significantly the predictions obtained with a parabolic confining potential. There is only one difference: the dispersion curves of different sublevels, calculated with the triangular well model, are not only diamagnetically shifted in energy with respect to the $B = 0$ case but have also their minima shifted with respect to $k_x$ since the energy depends in this case on the position of the orbit center (as follows from Eq.(3) the electron energy does not depend on the position of the orbit center $\bar{z}$ for a parabolic confining potential). Therefore, the magnetic depopulation in 2DEG confined by triangular wells would be accompanied also by momentum jumps of electrons, but the new directions of electron propagation would be asymmetric with respect to the initial direction of electron propagation chosen also as direction of the a[plied magnetic field.

The momentum jumps appear only when the magnetic field is in the plane of the 2DEG gas. Otherwise, for a magnetic field along the $z$ direction, the electron propagation in the 2DEG plane is constrained to Landau orbits. The energy spectrum of electrons is discrete, so that no momentum jump can occur, the magnetic field depopulation being associated to Shubnikov-de Hass oscillations or to the quantum Hall effect.

This momentum jump can be used, for example, as a switch for electrons or as an implementation of a qubit in quantum computers based on ballistic electrons. If initially more than two subbands would been occupied with electrons, several momentum jumps would occur for different values of the perpendicular magnetic field, at each jump two new electron propagation direction appearing, symmetric with respect to the direction of $B$ (if we consider a symmetric, parabolic confining potential $V(z)$ ). The new directions of electron propagation can be made asymmetric with respect to $B$ if the electrons are launched in the parabolically confined 2DEG such that initially $k_x \neq 0$. The new directions of electron propagation are asymmetric, even when initially $k_x = 0$, if the confining potential is not symmetric with respect to $z$, e.g. it has a triangular shape.

Can these jumps in ballistic electron momentum be observed? To answer this question let us consider a parabolically confined 2DEG with $N_1 = 1.2 \times 10^{14}$ m$^{-2}$ in the absence of the magnetic field, $m = 0.067\, m_0$ where $m_0$ is the free electron mass and $z_0 = 4$ nm. Then, the magnetic field necessary to depopulate this subband, calculated according to Eq.(6), is $B_1 = 6.4$ T, a value comparable to those reported in Ref.[2]. For this value of the applied magnetic field momentum jumps appear at $\boldsymbol{q} = \pm 10°$ if the electrons propagate initially in the $y$ direction with a mobility $\boldsymbol{m} = 6 \times 10^2$ m$^2$V$^{-1}$s$^{-1}$ under the action of an electric field of only $F = 640$ V/m. A larger electric field reduces the deviation angle. For this value of the critical magnetic field the electron wavefunction is not significantly perturbed with respect to the $B = 0$ case since $\bar{z} / z_0 = 0.02$.

The value of the critical magnetic field is strongly dependent on the subband population $N_1$ and on the confinement potential in the absence of the magnetic field. For example, for the same value of $N_1$ and of the other parameters as before, a slightly less confined 2DEG, with $z_0 = 5$ nm, requires a parallel magnetic field of only $B_1 = 5.12$ T to depopulate the higher subband, and an application of an electric field of $F = 644$ V/m to obtain momentum jumps in the same $\boldsymbol{q} = \pm 10°$ directions. The ratio $\bar{z}/z_0$ is in this case 0.038.

A $10°$ deviation from the initial electron trajectory is certainly observable; it can be demonstrated by collecting the electrons from the electrostatic gates placed in the positions indicated in Fig.2. In the experimental setup represented in Fig.2 E and C denote the electron emitter and collector, defined by electrostatic gates; an electric field E must be applied between them to force the electrons to propagate along the solid line at low magnetic fields. The width of the constriction in E should be sufficiently large such that the electron beam is not divergent. Detectors $D_1$ and $D_2$ collect the electrons that deviate from their initial direction of propagation when the critical magnetic field $B_1$ is attained; the trajectories of these electrons are represented with dashed line.

A larger deviation is obtained for a small $k_y$, i.e. for a small value of the electric field $F$. The experimental evidence of the momentum jumps are within the present technological possibilities: 2DEG ballistic electron propagation in modulation-doped GaAs/AlGaAs heterojunctions with parameters similar to those used in the simulations have been already demonstrated over elastic mean free paths of more than 60 µm at a temperature of 0.3 K [7]. The fact that momentum jumps have not been observed experimentally by now is not surprising. The main experimental effort up to now has been focused on magnetoresistance measurements, where the quality of the 2DEG samples are less restrictive than in electron optics experiments reported for example in Ref.[7] (see also Ref.[8] and the references therein). On the other hand, electron optics experiments are not usually performed at high magnetic fields. Moreover, the theory that modeled the result of the previous magnetoresistance measurements has not assumed ballistic (collisionless) electron propagation since 2DEG of the required quality were not available of that time. Thus it is

9understandable why momentum jumps have not been theoretically or experimentally evidenced up to now.

The presence or absence of electrons at detectors $D_1$ or $D_2$ (see Fig.) can constitute two logic states of a qubit. Since after the momentum jump there are in fact three electron beams, one could in principle associate this beams to three logic states of a qubit in a multi-value logic but this possibility is hampered by the fact that the three logic states (in particular, the outer electron beams) cannot be controlled independently. Another possible application of these momentum jumps is an electron switch: there is or not current at detectors $D_1$ and $D_2$ depending on the value of the magnetic field applied parallel to the 2DEG plane.

In conclusion, a new phenomenon has been predicted for ballistic electrons in quasi-2DEG systems. Momentum jumps of such electrons appear whenever higher magnetoelectric subbands are depopulated in the presence of an increasing magnetic field applied parallel to the 2DEG plane. The momentum jumps create new propagation directions for electrons, which may or not be symmetric with respect to the initial electron trajectory depending whether the confinement potential in the absence of the magnetic field is or not symmetric and whether the electron trajectories are initially parallel or not to the direction of the magnetic field. No such jumps are expected to exist for magnetic fields applied perpendicular to the 2DEG since the electrons are confined by the electrostatic potential in the direction of the magnetic field and also confined in Landau orbits in the direction transverse to the magnetic field. This phenomenon can be demonstrated experimentally and can be used to implement quantum logic states or electron switches.

10REFERENCES[1]    D.K. Ferry and S.M. Goodnick, *Transport in Nanostructures* (Cambridge Univ. Press, Cambridge, 1997).[2]    D.J. Newson, K.-F. Berggren, M. Pepper, H.W. Myron, G.J. Davies, and E.G. Scott, J. Phys. C **19**, L403 (1986).[3]    R.E. Doezema, M. Nealon, and S. Whitmore, Phys. Rev. Lett. **45**, 1593 (1980).[4]    M.A. Brummell, M.A. Hopkins, R.J. Nicholas, J.C. Portal, K.Y. Cheng, and A.Y. Cho, J. Phys. C **19**, L107 (1986).[5]    D.A. Wharam, T.J. Thornton, R. Newbury, M. Pepper, H. Ahmed, J.E.F. Frost, D.G. Hasko, D.C. Peacock, D.A. Ritchie, and G.A.C. Jones, J. Phys. C **21**, L209 (1988).[6]    J.A. Lebens, R.H. Silsbee, and S.L. Wright, Phys. Rev. B **37**, 10308 (1988).[7]    J. Spector, H.L. Stormer, K.W. Baldwin, L.N. Pfeiffer, and K.W. West, Appl. Phys. Lett. **56**, 1290 (1990).[8]    D. Dragoman and M. Dragoman, *Quantum-Classical Analogies* (Springer, Berlin, 2004).

FIGURE CAPTIONS

Fig.1 (a) Geometry of the ballistic quasi-2DEG system. The electron trajectories are depicted with solid line for $B < B_1$ and with dashed line for $B > B_1$. (b) Energy dispersion diagram of the two lowest subbands, depicting the electron jumps at $B_1$.

Fig.2 Experimental setup to demonstrate the momentum jumps

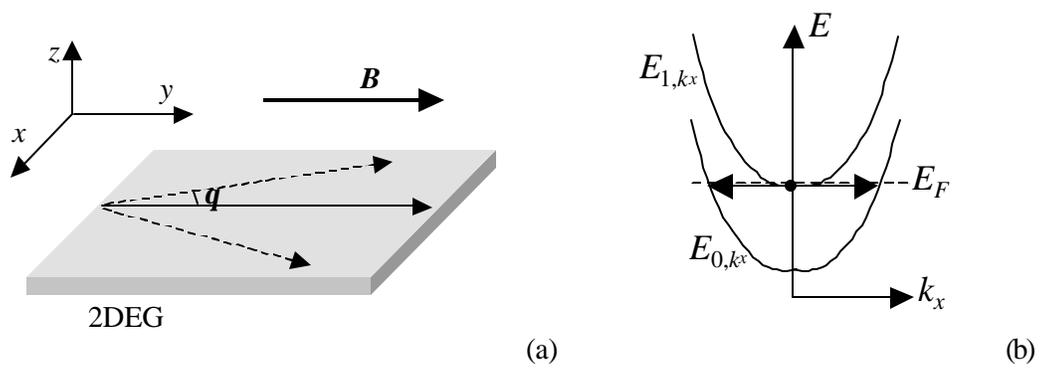

Fig.1

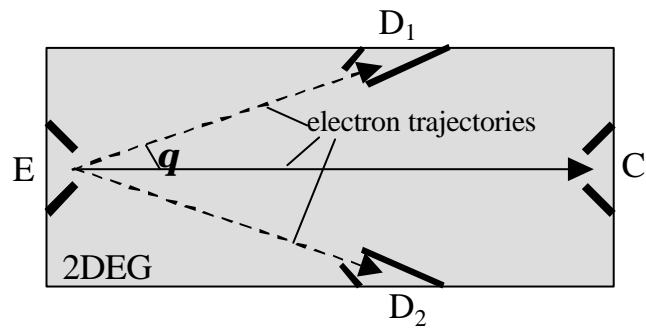

Fig.2